\documentclass[runningheads]{llncs}
\usepackage[T1]{fontenc}
\usepackage{graphicx}
\usepackage{braket}
\usepackage[cmex10]{amsmath}
\usepackage{xcolor}

\usepackage{changes}
\definechangesauthor[name={Ernesto Acosta}, color=green]{EA}
\definechangesauthor[name={Guillermo Botella}, color=red]{GB}
\definechangesauthor[name={Carlos Cano}, color=brown]{CC}

\usepackage{hyperref}

\begin{document}
\title{Adiabatic training for Variational Quantum Algorithms}
%
%
\author{Ernesto Acosta\inst{1}\orcidID{0009-0001-4911-6819}  \and
Carlos Cano Gutiérrez\inst{1}\orcidID{0000-0002-0181-2444} \and
Guillermo Botella\inst{2}\orcidID{0000-0002-0848-2636} \and
Roberto Campos \inst{3}\orcidID{0000-0002-2527-4177}}
\authorrunning{Ernesto Acosta, Carlos Cano, Guillermo Botella, Roberto Campos.}
%
\institute{Dpt. Computer Science and AI. University of Granada. 18071 Granada, Spain. \email{eacostam@correo.ugr.es} \and
Computer Architecture and Automation Department, Complutense University of Madrid, 28040 Madrid, Spain\\
\and
Theoretical Physics Department, Complutense University of Madrid, 28040 Madrid, Spain\\
}

%
\maketitle              

\begin{abstract}
This paper presents a new hybrid Quantum Machine Learning (QML) model composed of three elements: a classical computer in charge of the data preparation and interpretation; a Gate-based Quantum Computer running the Variational Quantum Algorithm (VQA) representing the Quantum Neural Network (QNN); and an adiabatic Quantum Computer where the optimization function is executed to find the best parameters for the VQA.

As of the moment of this writing, the majority of QNNs are being trained using gradient-based classical optimizers having to deal with the barren-plateau effect~\cite{ref_article_cerezo2021}. Some gradient-free classical approaches such as Evolutionary Algorithms have also been proposed to overcome this effect ~\cite{ref_article_acampora}. However, to the knowledge of the authors, adiabatic quantum models have not been used to train VQAs.

The paper compares the results of gradient-based classical algorithms against adiabatic optimizers showing the feasibility of integration for gate-based and adiabatic quantum computing models, opening the door to modern hybrid QML approaches for High Performance Computing. 

\end{abstract}

\section{Introduction}
Quantum Machine Learning has emerged during the last decade~\cite{ref_book_schuld2021ml} as a discipline in itself, at the intersection between (classical) Artificial Intelligence and Quantum Computing, in an attempt to extend the frontier of AI capabilities to the next level by taking advantage of the properties of quantum mechanics such as superposition, entanglement, tunnel effect, among others.  However, it is important to highlight that, being a novel technology, it might go through a stage of low expectations and general interest, in what would be called \textit{Quantum Winter}, analog to the well-known \textit{winters of Artificial Intelligence}~\cite{ref_book_russel2010}.

One of the areas of big research interest is the financial sector, particularly on the Time Series prediction as a mechanism to anticipate future market fluctuations.  For this purpose, a large number of mathematical and statistical models have emerged.  It has been, thanks to AI, that new machine learning models such as Artificial Neural Networks (ANNs), have made a great contribution to these research efforts, with Recurrent Neural Networks(RNNs) being one of the most widely accepted models~\cite{ref_book_werbos1975}.

This paper presents a Quantum Computing ~\cite{ref_book_nielsen} 
model approach based on training Variational Quantum Algorithms (VQA) by means of Adiabatic Quantum Computing~\cite{ref_article_farhi} for the analysis of the stock price trend for a given listed company.
To date, as far as we know, Adiabatic Quantum Computers have not been used to training VQAs. The closest research available to date has been proposed by Zhao and Xiao in 2021~\cite{ref_article_zhao2021} although their work deals only with binarized neural networks.

\section{Quantum RNN}
One of the most widespread Quantum Machine Learning models at present are the Variational Quantum Algorithms~\cite{ref_article_cerezo2021}, which are arguably the quantum equivalents to ANNs. Those are made up of Quantum Gates whose rotation angles have to be adjusted the same way as the neuron connection weights in ANNs.

For Time Series research, a variety of classical models have been defined (see, for example~\cite{ref_article_imf_forecasting}, ~\cite{ref_article_kantidakis_2022}, ~\cite{ref_article_khalique_2022}), among which RNNs became very popular because of their learning capability~\cite{ref_article_rumelhart_1986}.
The Quantum Recurrent Neural Network (QRNN) model of reference for this work is the one defined by Yanan Li~\cite{ref_article_li2023quantum}. The network is designed by  sequentially concatenating a circuit called Quantum Recurrent Block (QRB), as many times as the historical depth of the time series used for  prediction. This model proposes the search for the optimal parameters through gradient-based optimizers for a given cost function~\cite{ref_article_li2023quantum}.

In this paper, we explore how to translate the problem of finding the optimal parameters of a QNN into a QUBO optimization problem, and its resolution by means of Adiabatic Quantum Computation.

\subsection{Quantum Recurrent Block (QRB)}
Equivalent to a classical RNN, a Recurrent Block receives an input value $x^{(t)}$, holds an internal state value $h^{(t-1)}$, and produces an output value $y^{(t)}$.

With the sequential arrangement of several of this Recurrent Blocks, the network is able to read, in turns, various historical values and keep acquired knowledge inside the network by propagating the internal state values $h$, generating an output value $y^{(t+1)}$ after the last block.  By this, each Recurrent Block will have two independent inputs: the external data $x^{(t)}$ and the previous block's internal state $h^{(t-1)}$.

\begin{figure}
\setlength{\belowcaptionskip}{-20pt}
\centering
\includegraphics[width=0.6\textwidth]{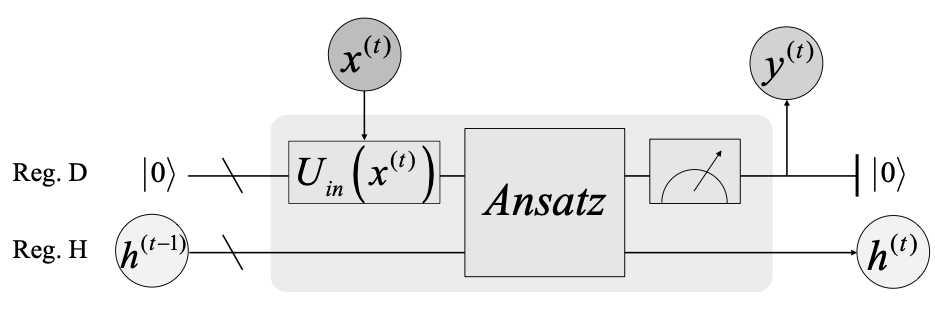}
\caption{Quantum Recurrent Block (QRB). Source:~\cite{ref_article_li2023quantum}}
\label{qrb}
\end{figure}

In general, the QRB is composed of three parts: input data encoding $Uin(x)$, Ansatz circuit and partial measurement.  The qubits are grouped in two registers: Register D, which encodes the input values, and Register H, which stores the historical information learned from the previous steps (see Figure \ref{qrb}). 

\subsubsection{Input data encoding.}
This process requires the application of a unitary transformation $Uin(x)$ to the input vector containing classical data.  That is $\ket{f(x)} = Uin\ket{0}^{\otimes n}$ where \textit{n} is the number of qubits.  In QNNs it is common to use angle encoding techniques, where a classical data is represented by the rotation of angles of one qubit, or controlled rotations in two-qubit gates.

In this work, we use the original encoding scheme consisting in the application of rotation on the \textit{x} axis, where the angle of rotation is the arc cosine of the normalized input value $\tilde{x}$.  The input value is replicated in several input qubits of Register D as this has shown to provide higher accuracy ~\cite{ref_article_gil2020}.

\subsubsection{Ansatz subcircuit.}
This is a parametric circuit composed of a polynomial number of quantum gates, whose rotation angles are optimized in the training process. As in ANNs, the universal approximation theorem defines that within a certain margin of error there always exists a quantum circuit that can approximate an objective function~\cite{ref_article_cai2022sample}.

There are three fundamental Ansatz configurations based on two-qubit gates: NN (Nearest Neighbour), AA (All to all) and CB (Circuit block). We chose the CB configuration for this work, as it offers a good balance between expressiveness and circuit size~\cite{ref_article_li2023quantum}.

\begin{figure}[hbtp]
\setlength{\belowcaptionskip}{-10pt}
\centering
\includegraphics[width=0.8\textwidth]{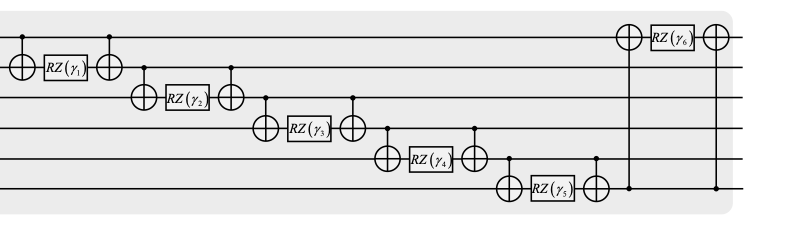}
\caption{Ansatz subcircuit design. Source:~\cite{ref_article_li2023quantum}}
\label{ansatz_final}
\end{figure}

The final Ansatz circuit involves \emph{Rzz($\theta$)} gates, which can be implemented by CNOT and Pauli-Z gates (see Figure \ref{ansatz_final}). 

\subsubsection{Partial Measurement.}
In the end of each quantum block a Z-axis measurement is performed on the first qubit of  Register D. The probability of this first qubit collapsing to state $\ket{1}$ is taken as prediction of $y_t$. After this, all qubits in Register D are reset to state $\ket{0}$ so they can encode the next input value.

\subsection{Recurrent Blocks Arrangement}
\begin{figure}[hbtp]
\setlength{\belowcaptionskip}{-20pt}
\centering
\includegraphics[width=1\textwidth]{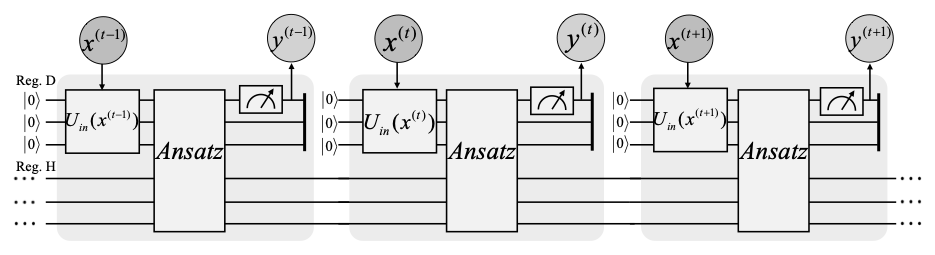}
\caption{pQRNN architecture. Source:~\cite{ref_article_li2023quantum}}
\label{pqrnn}
\end{figure}

We chose the Plain QRNN (pQRNN) architecture~\cite{ref_article_li2023quantum}, consisting of a strictly sequential arrangement where the D and H registers maintain their initial function throughout the network (see Figure \ref{pqrnn}).  The measurement qubit is in Register D while Register H maintains learning internal state and remains unmeasured through the circuit. Therefore Register H must keep coherence during the whole execution, which is a decisive aspect in current NISQ devices.

\section{Adiabatic Training for QRNNs}

In Adiabatic Quantum Computing, the evolution of the Hamiltonian in time \emph{H(t)} changes the state of the system according to the Schrödinger equation:

\begin{equation}
H(t) \ket{\psi(t)} = i\hbar \frac{\partial}{\partial t} \ket{\psi(t)}
\end{equation}

\subsubsection{Quantum Annealing.}
Quantum annealing is one way to perform an Adiabatic process by using an external magnetic field playing the same role as the temperature in the simulated annealing process. Thus, Quantum Annealing starts with a high magnetic field and ends with a low magnetic field. The \emph{Quantum Adiabatic Algorithm} is used to identify that the system is in the minimum energy state by defining a Hamiltonian \emph{Ising}.

The Ising model describes the behavior of a set of spins $z_i$ that interact with each other and with external forces $h_i$. Each spin has two configurations: $z_i \in \{-1,1\}$, such that the energy is defined by the following Hamiltonian:

\begin{equation}
- \sum_{j,k} J_{jk}z_{j}z_{k} - \sum_{j}h_{j}z_{j}
\end{equation}

\subsubsection{QUBO formulation.}
One way to implement the \emph{Quantum Adiabatic Algorithm} is by the QUBO (Quadratic Unconstrained Binary Optimization) formulation:
\begin{equation}
Q(x) = \sum_{i} a_{i}x_{i} + \sum_{ij} b_{ij}x_{i}x_{j}
\end{equation}

\begin{itemize}
    \item $x_i$ are binary variables that can take values of 0 or 1.
    \item $a_i$ represent the linear coefficients associated with binary variables $x_i$
    \item $b_{ij}$ are the quadratic coefficients associated with pairs of bin variables $x_i$, $x_j$
\end{itemize}

\subsubsection{Proposed QRNN optimization model.}
Therefore, we accomplished the search for optimal parameters of the Ansatz circuit by defining an optimization problem in a QUBO formulation. For this, we need to compute the Quantum Operator of this circuit and use it in the minimization of the error between the QRNN predicted and the expected values, as detailed in the following section.

\section{Implementation}

\subsection{Problem and data description}
Given the historical closing stock price values of a listed company, the problem is to find the trend (increase / decrease) for the next closing price.

In order to process historical data, the training dataset must be prepared containing the last two values ($x(t), x(t-1)$) along with the closing indicator (\emph{increment/decrement, ie. 1/0}) for each trading day as shown in Table \ref{table:dataset_training}. 

    \begin{table}[htbp]
        \setlength{\belowcaptionskip}{-20pt}
        \centering
        \begin{tabular}{||c c c c||} 
         \hline
         x(t) & x(t-1) & x(t-2) & y(t) \\ [0.5ex] 
         \hline\hline
         0.185520 & 0.209653 & 0.209653 & 0 \\ 
         0.209653 & 0.125189 & 0.125189 & 1 \\ 
         0.125189 & 0.444947 & 0.444947 & 0 \\ 
         0.444947 & 0.143288 & 0.143288 & 1 \\ 
         0.143288 & 0.319759 & 0.412647 & 0 \\ 
         ... & ... & ... & ... \\ [1ex] 
         \hline
        \end{tabular}
        \caption{Training dataset.  History depth: 3 days}
        \label{table:dataset_training}
    \end{table}

\subsection{Architecture of the proposed solution}
\begin{figure}[h!]
    \centering
    \includegraphics[width=1\textwidth]{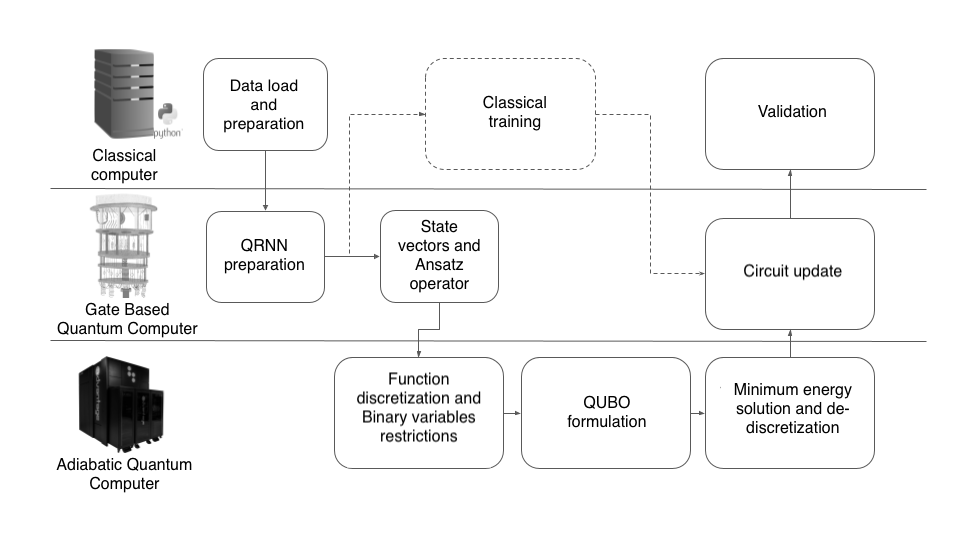}
    \caption{Execution Flow.}
    \label{execution_flow}
\end{figure}

The workflow diagram in Figure \ref{execution_flow} depicts the overall process proposed, including both the classical training and the proposed adiabatic quantum model for learning the parameters for the VQAs. \footnote{The implementation code is available in the github repository:~\url{https://github.com/eacostam/adiabaticVQA.git}.}

\begin{figure}[h!]
    \centering
    \includegraphics[width=1\textwidth]{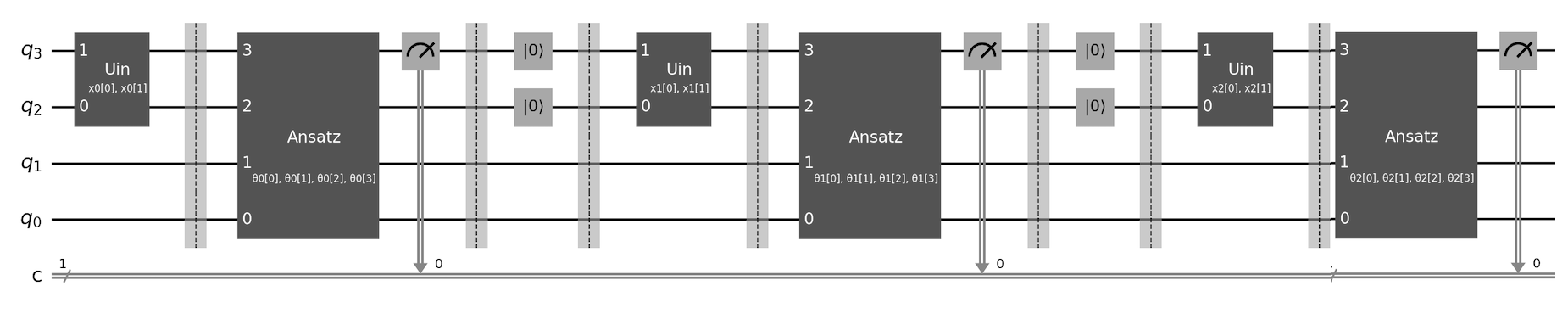}
    \caption{Quantum Recurrent Neural Network containing three QRB, with identical ansatz, processing three historical values, each value duplicated in $q_2$ and $q_3$.}
    \label{circuit_full}
\end{figure}

For the Quantum Neural Network, two or more recurrent blocks are concatenated, obtaining a VQA as in Figure \ref{circuit_full}.The first qubit in Register D is partially measured at the end of each block and then all qubits in Register D are collapsed to $\ket{0}$ state.  The trend value is calculated by measuring the last qubit from Register D by the end of the circuit.

\subsection{Classical Training}
The clasical training of the QRNN is done using the traditional ANN mechanisms by means of gradient descent algorithms, where the difference between predicted and expected output values is calculated iteratively for each of the records of the data set using Simultaneous Perturbation Stochastic Approximation (SPSA) optimizer.  

The objective function is defined as follows:

\begin{equation}
\frac{1}{N} \sum_{p} \log (p + e^{-10})
\end{equation}

Once the training is complete, the average regression accuracy (hit counts) is calculated to evaluate the model.

\subsection{Adiabatic Quantum Training}
An optimization problem defined as a QUBO formulation is addressed towards minimizing the error between predicted $\tilde{y_t}$ and expected output values $y_t$, by using the Mean Square Error (MSE) metric. 

A proper representation of the normalized input $\tilde{x_t}$ and predicted output value $y_t$ is required for this QUBO formulation, as well as the matrix representation of the Ansatz Quantum Operator $Q(x, \theta)$.

\subsubsection{State Vectors.}
The vector representation is obtained for the input values $x_t$ (encoded by the \emph{Uin} subcircuit) as well as the expected output values (after partial measurement).

For instance, input value $x = 0.58$\footnote{10 digit decimal precision were used in the experiments, rounded to 2 digits in this writing for better readability.} will have the following state vector, after applying $R_y(\arccos(0.58))$ on each qubit in \textit{Register D}, while having \textit{Register H} qubits all in status $\ket{0}$.  The wave function will only have non-zero values for the states $\ket{0000}$, $\ket{0100}$, $\ket{1000}$ and $\ket{1100}$, which are components 0, 4, 8 and 12 in the state vector:

\begin{equation}
\ket{\psi_{x0}} = (0.79, 0, 0, 0, 0.41, 0, 0, 0, 0.41, 0, 0, 0, 0.21, 0, 0, 0)
\end{equation}

Similarly, measuring a value $\ket{1}$ in the first qubit or Register D will have the following state vector representation:

\begin{equation}
\ket{\psi_{y0}} = (0, 0, 0, 0, 0, 0, 0, 0, 1, 0, 0, 0, 0, 0, 0, 0)
\end{equation}

Note that for adiabatic optimization, we can focus only on the $0$ and $8$ coefficients of the output state vector, given that all other qubits must be zero or close to zero as those are not being measured.

\subsubsection{Ansatz Operator.}
The quantum operator corresponding to the Ansatz subcircuit is a 16x16 matrix of the following form:

\begin{center} 
\small
    \setcounter{MaxMatrixCols}{4}
    \setlength{\arraycolsep}{2pt}
    $\begin{pmatrix}
    e^{-i\theta_0 /2}e^{-i\theta_1 /2}e^{-i\theta_2 /2}e^{-i\theta_3 /2} & 0 & \cdots & \cdots \\
    0 & e^{-i\theta_0 /2}e^{-i\theta_1 /2}e^{i\theta_2 /2}e^{i\theta_3 /2} & \cdots & \cdots \\
    \cdots & \cdots & \cdots & \cdots \\
    \cdots & 0 & \cdots & e^{-i\theta_0 /2}e^{-i\theta_1 /2}e^{-i\theta_2 /2}e^{-i\theta_3 /2} \\
    \end{pmatrix}$
\end{center}

\vspace{10pt}
This is a diagonal matrix with the following elements in its diagonal (for ease of reading, only positions $0$, $1$ and $15$ are presented): 

\vspace{8pt}
$[e^{i\frac{-\theta_0-\theta_1-\theta_2-\theta_3}{2}},
e^{i\frac{-\theta_0-\theta_1+\theta_2+\theta_3}{2}}, \cdots ,e^{i\frac{-\theta_0-\theta_1-\theta_2-\theta_3}{2}}
]$

\subsection{Discretization of rotation angles}
\begin{figure}[htbp]
    \setlength{\belowcaptionskip}{-20pt}
    \centering
    \includegraphics[scale=1]{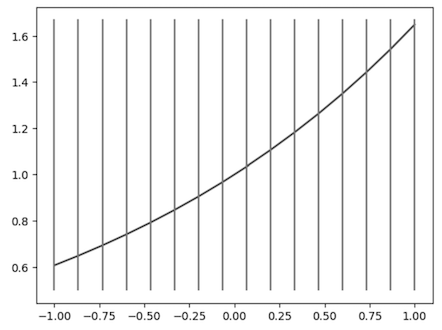}
    \caption{Discretization of the exponential function.}
    \label{discrete_exp}
\end{figure}

Adiabatic model uses binary variables and the Ansatz requires real values for the $\theta$ angles in the range $[0, \pi]$ for the exponential function, it is needed to discretize those possible angle values, hence loosing precision (see Fig. \ref{discrete_exp}).

\subsubsection{Ansatz Operator.}
The ansatz operator is applied to each input vector via dot product. Here we omit the phase factor \emph{i} given that the input state vectors have no imaginary component in any of their coefficients, and the QUBO formulations require only real values.

\subsubsection{QUBO Formulation.}
The QUBO formulation yields an optimization based on least squares differences, which, in the case of the QRNN, involves the sum of squares differences between the expected ($y_t$) and predicted ($\tilde{y_t}$) values only for coefficients $0$ and $8$ of the output state vectors:

\begin{equation}
L_2(\vec{\theta})=\frac{1}{N} \sum_{t=0}^{n-1}\left(\tilde{y_t}[0](x_t,\vec{\theta})-y_t[0]\right)^2 + \left(\tilde{y_t}[8](x_t,\vec{\theta})-y_t[8]\right)^2
\end{equation}

As a reminder, the circuit only measures the first qubit in \textit{Register D},  which is equivalent to having the entire system in either status $\ket{0000}$ or $\ket{1000}$, which correspond to coefficients 0 and 8 in the wave equation.

This way, the QUBO formulation is proposed, optimizing the difference by least squares over the set of training values $[(x_t, x_{t-1}), (y_t)]$. After the adiabatic process is complete, an inverse discretization is applied in order to obtain the real values corresponding to the identified $\theta$ angles.

\section{Results}
The following are the results obtained in a typical execution of the model, for the three scenarios: 1) QRNN with classical training, 2) QRNN with adiabatic training and 3) ANN, Classical Artificial Neural Network. The accuracy for the different models is shown in Table \ref{table:precisiones}.

All the executions where performed on a local personal computer on a python framework. Therefore, both gate-based and annealing quantum models were simulated using the corresponding software libraries from IBM-Qiskit and D-Wave Ocean. Validation of the obtained results against executions on real noisy hardware will be subject of further research. 

The hardware configuration used was: OS MS Windows 11 version 10.0.22621, 1 CPU Intel64 Family 6 Model 140 GenuineIntel 2.995Mhz, and Physical Memory 32,602MB.
The datasets used were obtained from Yahoo Finance website for selected listed companies mentioned in \ref{table:accuracy} were the closing stock price is registered for each trading day.  In order to perform cross validation, the dataset has been split in two parts: 80\% used for training, 20\% for testing.
\begin{table}[htbp]
    \setlength{\belowcaptionskip}{-10pt}
    \centering
    \begin{tabular}{||c c c c||} 
     \hline
         & QRNN & QRNN &  ANN\\ 
         & Classical Training & Adiabatic Training &  \\ 
         \hline
         \textbf{Accuracy} & \textbf{66\%} & \textbf{66\%} & 64\% \\  
     \hline
    \end{tabular}
    \caption{Model accuracy.}
    \label{table:precisiones}
\end{table}

Discretization is done in 4 parts, for simplicity, obtaining the values for the angles and the corresponding exponential function as seen in Table \ref{table:ent_adiabatico_angles}. 

\begin{table}[htbp]
    \setlength{\belowcaptionskip}{-10pt}
    \centering
    \begin{tabular}{||c c c c c ||}
        \hline
         & value 1 & value 2 & value 3 & value 4 \\ [0.5ex] 
        \hline
        \hline
        $\theta$ & 0,125 & 0,375 & 0,625 & 0,875 \\ [1ex]
        $exp(\theta /2)$ & 1,06449 & 1,20623 & 1,36683 & 1,54883 \\ [1ex]
        \hline
    \end{tabular}
    \caption{Discretized angles and corresponding exponential function values.}
    \label{table:ent_adiabatico_angles}
\end{table}
The total number of potential solutions explored was 1.048.576.
Once filtered as to keep only the feasible solutions (those that meet all the conditions of the QUBO formulation) 1.024 potential solutions were obtained.

The optimal solution found (the first in the list of feasible solutions) and its decoding into the real values of rotation angles is as follows:
$\theta_0$ = 0,625, $\theta_1$ = 0,375, $\theta_2$ = 0,625, $\theta_3$ = 0,375.

The QRNN is updated with these optimal values and validation is executed in the same way, obtaining 49 test records.  The obtained accuracy is 66\%.

\section{Results analysis}
\subsubsection{Accuracy.}

Table~\ref{table:accuracy} shows the accuracy of the proposed methods for predicting different stock market values with varying depth in the number of records.
Although accuracy is not much higher on adiabatic training, there is a slight positive difference with respect to classical training when the volume of data used for training is greater than 200 records.
These data show no significant difference in terms of accuracy between QRNN and ANN results, probably due to the high maturity of ANN models and the fact that it is a relatively small data set which leads to a quick convergence.

\begin{table}[hbtp]
    \setlength{\belowcaptionskip}{-20pt}
    \centering
    \begin{tabular}{||c c c c c||} 
     \hline
     Symbol & Number of & QRNN & QRNN & ANN \\ [0.5ex] 
      & records & Classical Training & Adiabatic Training &  \\ [0.5ex]
     \hline\hline
     SAN.MC & 250 & 65\% & \textbf{66\%} & 64\% \\  
     SAN.MC & 125 & 62\% & 59\% & \textbf{66\%} \\ 
     T & 100 & \textbf{57\%} & 47\% & 52\% \\
     SAN.MC & 150 & 64\% & \textbf{65\%} & 64\% \\
     RVPH & 250 & 68\% & 69\% & \textbf{71\%} \\ [1ex] 
     \hline
    \end{tabular}
    \caption{Accuracy for different large listed companies  (SAN.MC=Santander Bank, T= AT\&T, RVPH=Reviva Pharmaceuticals).}
    \label{table:accuracy}
\end{table}

\subsubsection{Execution counts (EC) and elapsed time}
Table~\ref{table:cycles} shows the execution counts and elapsed time in seconds for the proposed methods on the prediction of the stock market values scenarios in Table~\ref{table:accuracy}. Regarding QRNN with classical training, EC refer to epochs and circuit shots. In contrast, in adiabatic training, EC are adiabatic steps that depend on the discretization of the $exp(\theta)$ function rather than the number of records. In ANN, EC refer to epochs and records. 

EC and execution time are directly related to the volume of data being processed both for QRNN with classical training as well as in ANN as seen in Table \ref{table:cycles}. However, for adiabatic training, the EC is constant and the time required is substantially smaller for a fixed number of binary variables (see Table~\ref{table:timing_adiabatic}). This is due to the adiabatic process exploring the entire state space, in one single run until reaching the lowest energy state.  This process is dependent on the number of binary variables used rather than the number of records used by the optimization function.  The quantum annealing simulator gives no access to the details of the internal adiabatic evolution through time as to see the number of steps required to reach the lowest energy state. 

A special behavior is seen in the QRNN under classical training, where EC is greater than the number of iterations defined. This is due to the fact that the SPSA optimizer uses several stopping criteria, with the number of iterations being one of them but considering also the degree of convergence which might cause the additional iterations before full stop. The time and memory required for adiabatic training is directly related to the number of parts in which the exponential function is discretized. 

\begin{table}[hbtp]
    \setlength{\belowcaptionskip}{-5pt}
    \centering
    \begin{tabular}{||c|c|c|c|c||} 
     \hline
     Symbol & Number of & QRNN & QRNN & ANN \\ [0.8ex] 
      & records & Classical Training & Adiabatic Training &  \\ [0.8ex]
     \hline\hline
     SAN.MC & 250 & 900K (c) - 41K (s) & 1M (c) - 0,92 (s) & \textbf{25,0K (c)} - 14 (s) \\ 
     SAN.MC & 125 & 450K (c) - 22K (s) & 1M (c) - 0,92 (s) & \textbf{12,5K (c)} - 06 (s) \\ 
     T & 100 & 370K (c) - 13K (s) & 1M (c) - 1,62 (s) & \textbf{10,0K (c)} - 07 (s) \\
     SAN.MC & 150 & 496K (c) - 32K (s) & 1M (c) - 0,97 (s) & \textbf{15,0K (c)} - 11 (s) \\
     RVPH & 250 & 900K (c) - 41K (s) & 1M (c) - 0,93 (s) & \textbf{25,0K (c)} - 16 (s) \\ [1ex] 
     \hline
    \end{tabular}
    \caption{EC - Execution counts (c) and Elapsed time in seconds (s).}
    \label{table:cycles}
\end{table}

\begin{table}[htbp]
    \setlength{\belowcaptionskip}{-5pt}
    \centering
    \begin{tabular}{||c c c c c||} 
     \hline
     Parts & Binary & Explored & Feasible & Elapsed \\ [0.5ex] 
     & variables & options & options & time (s) \\ [0.5ex]
     \hline\hline
     3 & 12 & 4.096 & 81 & 0,926 \\ 
     4 & 16 & 65.536 & 256 & \textbf{0,92} \\ 
     5 & 20 & 1.048.576 & 625 & 0,927 \\ 
     6 & 24 & 16.777.216 & 1.296 & 1,13 \\ 
     \hline
    \end{tabular}
    \caption{Adiabatic optimization for SAN.MC(250) with discretization from 3 to 6.}
    \label{table:timing_adiabatic}
\end{table} 

\section{Conclusions}
Our results show that the accuracy of QRNN with adiabatic training is similar to QRNN with classical gradient-descend training. This poses an opportunity towards medium term quantum computers. Although it has not yet been demonstrated that Quantum Computers will be faster, it has been proven that their energy consumption will be smaller following C H. Bennett statement ~\cite{ref_article_bennet_1973}. 

The metrics obtained on this work were taken from simulated environments, therefore include an additional operational overhead.  Once the same models are run on real Quantum hardware, energy efficiency is expected to improve, making this approach a good candidate for High Performance Computing ecosystems.

As for the possibilities offered by Adiabatic Quantum Computing for this type of problems, the binary nature of the QUBO models makes the Variational Algorithms lose expressiveness because of the discretization of the rotation angles.  Exploration of alternatives to overcome this situation, like introduction of imaginary phase or iterative approximation methods, is recommended as future research works. Similarly, experimenting with more diverse datasets will help verify scalability and robustness of the proposed method.

At the same time, a new research path is presented towards trying to avoid the Barren Plateau effect thanks to the non-gradient dependency for the adiabatic optimization. A detailed comparison on this particular aspect is an interesting future research. As future work, it is suggested to explore alternatives in order to take advantage of the imaginary phase factor, and try an iterative approximation method within the vector space that might increase expressiveness of the model. 

\begin{credits}
\subsubsection{\ackname} 
This work was supported by Ecological and Digital Transition R\&D project  PID2021-128970OA-I00 call 2022 by MCIN/AEI/10.13039/501100011033 and European Union NextGeneration EU/PRTR. 
Additionally, this work received funding from grant PID2021-123041OB-I00 funded by MCIN/AEI/ 10.13039/501100011033 and by “ERDF A way of making Europe”. 
\subsubsection{\discintname}
The authors have no competing interests to declare that are
relevant to the content of this article. 
\end{credits}
%
%
%
%

\end{document}